\documentclass[10pt,final,journal,twocolumn]{IEEEtran}
\ifCLASSINFOpdf
\else
\fi
\usepackage{color}
\usepackage{cite}
\usepackage[cmex10]{amsmath}
\usepackage{amsthm}
\usepackage{algorithm}
\usepackage{algorithmic}
\usepackage{stfloats}
\usepackage{subfigure}
\usepackage{mathrsfs}
\usepackage{multicol,multienum}
\usepackage[dvips]{graphicx}

\begin{document}

\title{Unmanned Aerial Vehicle-Aided Communications: Joint Transmit Power and Trajectory Optimization}

\author{Haichao Wang,
        Guochun Ren,
        Jin Chen,
        Guoru Ding,
        and Yijun Yang

\thanks{This work is supported by the National Natual Science Foundation of China (Grant No. 61501510), Natural Science Foundation of Jiangsu Province (Grant No. BK20150717), China Postdoctoral Science Funded Project (Grant No. 2016M590398), and Jiangsu Planned Projects for Postdoctoral Research Funds (Grant No. 1501009A).}
\thanks{The authors are with the College of Communications Engineering, Army Engineering University of PLA, Nanjing 210007, China (e-mail: whcwl0919@sina.com, gc\underline{ }ren@sina.com, chenjin99@263.net, dr.guoru.ding@ieee.org, 13776601868yyj@sina.com). G. Ding is also with National Mobile Communications Research Laboratory, Southeast University, Nanjing 210096, China.}

}

\IEEEpeerreviewmaketitle
\maketitle
\begin{abstract}
This letter investigates the transmit power and trajectory optimization problem for unmanned aerial vehicle (UAV)-aided networks. Different from majority of the existing studies with fixed communication infrastructure, a dynamic scenario is considered where a flying UAV provides wireless services for multiple ground nodes simultaneously. To fully exploit the controllable channel variations provided by the UAV's mobility, the UAV's transmit power and trajectory are jointly optimized to maximize the minimum average throughput within a given time length. For the formulated non-convex optimization with power budget and trajectory constraints, this letter presents an efficient joint transmit power and trajectory optimization algorithm. Simulation results validate the effectiveness of the proposed algorithm and reveal that the optimized transmit power shows a water-filling characteristic in spatial domain.
\end{abstract}


\IEEEpeerreviewmaketitle

\section{Introduction}
\IEEEPARstart{U}{nmanned} aerial vehicles (UAVs) have attracted increasing attention recently since future applications claim for more autonomous and rapid deployable systems. Compared with communications with fixed infrastructure, UAV-aided networks bring additional gains with the inherent mobility\cite{ZRUAV,OurPaper}. To fulfill the UAVs' potentials, resource allocation for UAV-aided networks is crucial but challenging with the fact that the UAVs can move freely in the air.

The research on UAV-aided communications is still in its infancy. Most of existing studies focus on the efficient deployments of UAVs as mobile base stations (BSs)\cite{CL1,CL2,bsmm}, where the authors attempt to employ UAVs to provide wireless connectivity for ground users. However, UAVs are generally fixed in some places and thus the mobility is not fully utilized. In\cite{FT}, the authors investigate the UAV communication under hover time constraints considering a network in which multiple UAVs provide wireless service to ground users. To our best knowledge, the authors in\cite{relaytc} firstly investigate the power and trajectory optimization problem for a UAV-assisted mobile relay system, which shows that significant throughput gains can be achieved by exploiting the channel variations. Following\cite{relaytc}, authors in\cite{BSC} study a simplified case, where the users are equally spaced on the ground along a straight line. The previous observations motivate us to exploit the UAV's mobility to provide better service for randomly dispersed ground users by dynamically adjusting the UAV's locations and transmit power.

In this letter, we investigate the power allocation and trajectory optimization problem for UAV-aided networks, where a UAV provides network access for multiple nodes simultaneously. We formulate a non-convex optimization problem with the aim to maximize the minimum average throughput within a given time length, subject to the trajectory constraints and power budget. By exploiting the inherent characteristics of the formulated problem, we develop an efficient joint transmit power and trajectory optimization algorithm, where two subproblems are first investigated: Transmit power optimization with given trajectory and trajectory optimization with given transmit power. Moreover, a lower bound of the non-convex function in trajectory optimization is derived to address this subproblem. Simulation results validate the superiority of the proposed algorithm and reveal that the optimized transmit power shows a water-filling characteristic in spatial domain.
\section{System Model and Problem Formulation}
Consider a scenario where a set $\mathcal N= \left\{ {1,2,...,n,...,N} \right\}$ of $N$ nodes are randomly dispersed in the ground and a UAV flying at a fixed altitude $H$ provides network connectivity for these nodes within a finite time horizon $T$. Since the UAV's launching/landing locations are generally fixed for performing certain missions, the initial and final locations are given as $\left[ {{x_0},{y_0},{H}} \right]$ and $\left[ {{x_F},{y_F},{H}} \right]$, respectively. For convenience, denote $\left[ {{x_0},{y_0},{0}} \right]$ as the origin of the considered coordinate system. The total time length $T$ is divided into $M$ small time slots with each ${\delta}$ length, i.e., $T = M{\delta}$. Therefore, the UAV's trajectory can be approximated by $\left\{ {x\left[ m \right],y\left[ m \right],H} \right\}, m \in \mathcal M =\{1,...,M\}$. The number $M$ of discrete points makes a tradeoff between the computational complexity and the approximation accuracy. Specifically, larger number $M$ or smaller value ${\delta}$, on the one hand, results in much more optimization variables, increasing the complexity; on the other hand, it provides more accurate trajectory. Considering that the UAV's maximum flight speed is limited by $V$, there should be constraints on the UAV's locations as follows
\begin{align}
\label{trajectory}
&{\left( {x\left[ 1 \right] - {x_0}} \right)^2} + {\left( {y\left[ 1 \right] - {y_0}} \right)^2} \le {\left( {V\delta } \right)^2},\nonumber\\
&{\left( {x\left[ m \right] - x\left[ {m - 1} \right]} \right)^2} + {\left( {y\left[ m \right] - y\left[ {m - 1} \right]} \right)^2} \le {\left( {V\delta } \right)^2},\nonumber\\
&{\left( {{x_F} - x\left[ M \right]} \right)^2} + {\left( {{y_F} - y\left[ M \right]} \right)^2} \le {\left( {V\delta } \right)^2}.
\end{align}
Since the UAV's height $H$ is fixed during the flight, without loss of generality, we focus on the $\left\{ {x\left[ m \right],y\left[ m \right]} \right\}$ in the following analysis. Specifically, a downlink orthogonal frequency division multiple access is considered. The total bandwidth and transmit power are denoted by $B$ and $P_T$, respectively. Equal bandwidth is allocated to each served node. The channel power gain between the UAV and the $n$-th node at the $m$-th time slot ${g_n}\left[ m \right]$ is dominated by line-of-sight and given by\cite{relaytc,BSC,zy2}
\begin{align}
\label{cmodel}
{g_n}\left[ m \right] = \frac{{{\beta _0}}}{{{{\left( {x\left[ m \right] - {x_n}} \right)}^2} + {{\left( {y\left[ m \right] - {y_n}} \right)}^2} + {H^2}}},
\end{align}
where $\beta_0$ is the channel power gain at the reference distance $d_0$ and $(x_n,y_n,0)$ is the coordinate of $n$-th node. It can be observed from (\ref{cmodel}) that the channel power gain monotonically decreases with an increasing altitude $H$. In this case, lowest altitude is expected since it achieves best channel conditions. Therefore, we do not consider the optimization of the UAV's altitude in this letter. The average throughput received at the $n$-th node within the time length $T$ is
\begin{align}
\label{averageth}
{R_n} = \frac{1}{T}\sum\limits_{m = 1}^M {\frac{B}{N}{{\log }_2}\left( {1 + \frac{{{p_n}\left[ m \right]{g_n}\left[ m \right]}}{{{B \mathord{\left/
 {\vphantom {B N}} \right.
 \kern-\nulldelimiterspace} N}{\sigma ^2}}}} \right)},
\end{align}
where ${p_n}\left[ m \right]$ is the UAV's transmit power for $n$-th node and $\sigma ^2$ is the noise power spectrum
density. To ensure that all the ground nodes have communication opportunities, which is different from the winners-take-all objective, maximizing the minimum average throughput is considered via allocating the transmit power and optimizing the UAV's trajectory. Mathematically, the investigated problem can be formulated as follows:
\begin{align}
\label{ProblemO}
&\mathop {\max }\limits_{\left\{ {x\left[ m \right],y\left[ m \right]} \right\},\left\{ {{p_n}\left[ m \right]} \right\}} \mathop {\min}\limits_{n} {R_n} \nonumber\\
&s.t.~C1: \sum\limits_{n = 1}^N {\sum\limits_{m = 1}^M {{p_n}\left[ m \right]} }  \le {P_T}, \nonumber\\
&~~~~~C2: {p_n}\left[ m \right] \geq 0, \forall n \in \mathcal N, \forall m \in \mathcal M,  \nonumber\\
&~~~~~C3:{\left( {x\left[ 1 \right] - {x_0}} \right)^2} + {\left( {y\left[ 1 \right] - {y_0}} \right)^2} \le {\left( {V\delta } \right)^2}, \nonumber\\
&~~~~~C4:{\left( {x\left[ m \right] - x\left[ {m - 1} \right]} \right)^2} \nonumber\\
 &~~~~~~~~~~~~+ {\left( {y\left[ m \right] - y\left[ {m - 1} \right]} \right)^2} \le {\left( {V\delta } \right)^2},m = 2,...,M,\nonumber\\
&~~~~~C5:{\left( {{x_F} - x\left[ M \right]} \right)^2} + {\left( {{y_F} - y\left[ M \right]} \right)^2} \le {\left( {V\delta } \right)^2}.
\end{align}
The constraints $C1$ and $C2$ are power budget. $C3-C5$ are the location constraints introduced in (\ref{trajectory}). This is a non-convex optimization problem due to the coupling of transmit power and trajectory, which is intractable to be solved with standard convex optimization techniques.
\section{Joint Transmit Power and Trajectory Optimization}
By introducing a variable $s$, the original problem (\ref{ProblemO}) can be reformulated as follows\cite{BSC,MMIN1}
 \begin{align}
\label{Problem}
&\mathop {\max }\limits_{\left\{ {x\left[ m \right],y\left[ m \right]} \right\},\left\{ {{p_n}\left[ m \right]} \right\},s} s \nonumber\\
&s.t.~ {R_n} \ge s, \forall n \in \mathcal N,\nonumber\\
&~~~~~ C1-C5 ~\text{in}~ (4),
\end{align}
which is still non-convex. However, it can be observed that ${R_n}$ is concave about the transmit power $\left\{ {{p_n}\left[ m \right]} \right\}$ with given ${g_n}\left[ m \right]$. Moreover, a lower bound of ${R_n}$ can be found with the given transmit power. Based on these observations, two subproblems are first investigated: Transmit power optimization with given trajectory and trajectory optimization with given transmit power. Then, a joint transmit power and trajectory optimization algorithm is designed.
\subsection{Transmit Power Optimization with Given Trajectory}
Serving ground nodes can be triggered by a third party when UAVs are planned for some specific applications and services, such as aerial photographs and goods transportations. Thus, the trajectory is given in this case. With given trajectory ${\left\{ {x\left[ m \right],y\left[ m \right]} \right\}}, m=1,...,M$, the transmit power optimization problem is given as follows:
\begin{align}
\label{P11}
&\mathop {\max }\limits_{\left\{ {{p_n}\left[ m \right]} \right\},s} s\nonumber\\
&s.t.~C1:\frac{1}{T}\sum\limits_{m = 1}^M {\frac{B}{N}{{\log }_2}\left( {1 + \frac{{N{p_n}\left[ m \right]{g_n}\left[ m \right]}}{{B{\sigma ^2}}}} \right)}  \ge s,\forall n \in \mathcal N, \nonumber\\
&~~~~~C2:\sum\limits_{n = 1}^N {\sum\limits_{m = 1}^M {{p_n}\left[ m \right]} }  \le {P_T},\nonumber\\
&~~~~~C3: {p_n}\left[ m \right] \geq 0, \forall n \in \mathcal N, \forall m \in \mathcal M.
\end{align}
It is a standard convex optimization problem, and some existing algorithms can be used\cite{Convex}, such as the interior point method with the complexity of $O\left( {{N^3}{M^3}} \right)$. Moreover, a low complexity algorithm can be developed by following\cite{FIP}.
\subsection{Trajectory Optimization with Given Transmit Power}
Due to the UAV's hardware limitations, the transmit power may be given or fixed. With given transmit power $\left\{ {{p_n}\left[ m \right]} \right\}$, the trajectory optimization problem can be reformulated as follows:
\begin{align}
\label{p21}
&\mathop {\max }\limits_{\left\{ {x\left[ m \right],y\left[ m \right]} \right\},s} s \nonumber\\
&s.t.~C1: \begin{array}{l}
\frac{1}{T}\sum\limits_{n = 1}^N {\frac{B}{N}{{\log }_2}\left( {1 + \frac{{N{p_n}\left[ m \right]}}{{B{\sigma ^2}}}} \right.} \\
\left. {\frac{{{\beta _0}}}{{{{\left( {x\left[ m \right] - {x_n}} \right)}^2} + {{\left( {y\left[ m \right] - {y_n}} \right)}^2} + {H^2}}}} \right) \ge s
\end{array}, \forall n \in \mathcal N, \nonumber\\
&~~~~~C2:{\left( {x\left[ 1 \right] - {x_0}} \right)^2} + {\left( {y\left[ 1 \right] - {y_0}} \right)^2} \le {\left( {V\delta } \right)^2}, \nonumber\\
&~~~~~C3:{\left( {x\left[ m \right] - x\left[ {m - 1} \right]} \right)^2} \nonumber\\
 &~~~~~~~~~~~~+ {\left( {y\left[ m \right] - y\left[ {m - 1} \right]} \right)^2} \le {\left( {V\delta } \right)^2},m = 2,...,M,\nonumber\\
&~~~~~C4:{\left( {{x_F} - x\left[ M \right]} \right)^2} + {\left( {{y_F} - y\left[ M \right]} \right)^2} \le {\left( {V\delta } \right)^2},
\end{align}
where the constraint $C1$ is non-convex. To this end, following\cite{relaytc}, an efficient algorithm is developed by iteratively optimizing the objective with the lower bound of constraint $C1$.

Denote $\left\{ {{x^k}\left[ m \right],{y^k}\left[ m \right]} \right\}$ as the trajectory at $k$-th iteration, then the trajectory at $k+1$-th is given by $\left\{ {{x^{k + 1}}\left[ m \right],{y^{k + 1}}\left[ m \right]} \right\}$ with ${x^{k + 1}}\left[ m \right] = {x^k}\left[ m \right] + \Delta _x^k\left[ m \right]$ and ${y^{k + 1}}\left[ m \right] = {y^k}\left[ m \right] + \Delta _y^k\left[ m \right]$. $\Delta _x^k\left[ m \right]$ and $\Delta _y^k\left[ m \right]$ are the increments at the $k$-th iteration. Thus, we have $R_n^{k + 1} = {1 \mathord{\left/
 {\vphantom {1 T}} \right.
 \kern-\nulldelimiterspace} T}\sum\limits_{m = 1}^M {{B \mathord{\left/
 {\vphantom {B N}} \right.
 \kern-\nulldelimiterspace} N}r_{n,m}^{k + 1}} $ and
\begin{align}
&r_{n,m}^{k + 1} = {\log _2}\left( {1 + \gamma \frac{{{\beta _0}}}{{d_{n,m}^k + f\left( {\left\{ {\Delta _x^k\left[ m \right],\Delta _y^k\left[ m \right]} \right\}} \right)}}} \right),
\end{align}
where
\begin{align}
&\gamma  = {{N{p_n}\left[ m \right]} \mathord{\left/
 {\vphantom {{N{p_m}\left[ n \right]} {B{\sigma ^2}}}} \right.
 \kern-\nulldelimiterspace} {B{\sigma ^2}}},\nonumber\\
 &d_{n,m}^k = {\left( {{x^k}\left[ m \right] - {x_n}} \right)^2} + {\left( {{y^k}\left[ m \right] - {y_n}} \right)^2} + {H^2},\nonumber\\
&f\left( {\left\{ {\Delta _x^k\left[ m \right],\Delta _y^k\left[ m \right]} \right\}} \right) = \Delta _x^k{\left[ m \right]^2} + \Delta _y^k{\left[ m \right]^2}\nonumber\\
&~+ 2\left( {{x^k}\left[ m \right] - {x_n}} \right)\Delta _x^k\left[ m \right] + 2\left( {{y^k}\left[ m \right] - {y_n}} \right)\Delta _y^k\left[ m \right].
\end{align}
The operation ``$A$'' represents ${x^{k + 1}}\left[ m \right] = {x^k}\left[ m \right] + \Delta _x^k\left[ m \right]$ and ${y^{k + 1}}\left[ m \right] = {y^k}\left[ m \right] + \Delta _y^k\left[ m \right]$.
Since function ${\log _2}\left( {1 + {a \mathord{\left/
 {\vphantom {a {\left( {b + x} \right)}}} \right.
 \kern-\nulldelimiterspace} {\left( {b + x} \right)}}} \right)$ is convex, there is
\begin{align}
\label{Inequa}
{\log _2}\left( {1 + \frac{a}{{b + x}}} \right) \ge {\log _2}\left( {1 + \frac{a}{b}} \right) - \frac{a}{{\ln 2b\left( {a + b} \right)}}x,
\end{align}
which results from the first order condition of convex functions\cite{Convex}. Based on the inequality (\ref{Inequa}), we have\cite{relaytc,zy2}
\begin{align}
&r_{n,m}^{k + 1} \ge lbr_{n,m}^{k + 1} =  {\log _2}\left( {1 + \gamma \frac{{{\beta _0}}}{{d_{n,m}^k}}} \right) \nonumber\\
&~~- \frac{{\gamma {\beta _0}}}{{\ln 2d_{n,m}^k\left( {\gamma {\beta _0} + d_{n,m}^k} \right)}}f\left( {\left\{ {\Delta _x^k\left[ m \right],\Delta _y^k\left[ m \right]} \right\}} \right).
\end{align}
Given the trajectory $\left\{ {{x^k}\left[ m \right],{y^k}\left[ m \right]} \right\}$ at $k$-th iteration, the trajectory at $k+1$-th iteration can be obtained by solving the following optimization problem
\begin{align}
\label{P22}
&\mathop {\max }\limits_{\left\{ {\Delta _x^k\left[ m \right],\Delta _y^k\left[ m \right]} \right\},s} s\nonumber\\
&s.t.~C1:\frac{1}{T}\sum\limits_{m = 1}^M {\frac{B}{N}} lbr_{n,m}^{k + 1} \ge s,\forall n \in \mathcal {N},\nonumber\\
&~~~~~C2:{\left( {{x^k}\left[ 1 \right] + \Delta _x^k\left[ 1 \right] - {x_0}} \right)^2}\nonumber\\
 &~~~~~~~~~~~ + {\left( {{y^k}\left[ 1 \right] + \Delta _y^k\left[ 1 \right] - {y_0}} \right)^2} \le {\left( {V\delta } \right)^2},\nonumber\\
&~~~~~C3:{\left( {{x^k}\left[ n \right] + \Delta _x^k\left[ n \right] - {x^k}\left[ {n - 1} \right] - \Delta _x^k\left[ {n - 1} \right]} \right)^2}\nonumber\\
 &~~~~~~~~~~~+ {\left( {{y^k}\left[ n \right] + \Delta _y^k\left[ n \right] - {y^k}\left[ {n - 1} \right] - \Delta _y^k\left[ {n - 1} \right]} \right)^2}\nonumber\\
  &~~~~~~~~~~~  \le {\left( {V\delta } \right)^2},n = 2,...,N, \nonumber\\
&~~~~~C4:{\left( {{x_F} - {x^k}\left[ N \right] - \Delta _x^k\left[ N \right]} \right)^2}\nonumber\\
   &~~~~~~~~~~~+ {\left( {{y_F} - {y^k}\left[ N \right] - \Delta _y^k\left[ N \right]} \right)^2} \le {\left( {V\delta } \right)^2},
\end{align}
which is a convex optimization problem and can be solved using standard convex optimization techniques\cite{Convex}. Since the optimization variables are the increments at each iteration, a series of non-decreasing values can be obtained. On the other hand, these values must be upper bounded by the optimal solution to the problem (\ref{p21}). Therefore, the convergence is guaranteed.
\subsection{Joint Transmit Power and Trajectory Optimization}
Since the investigated joint trajectory optimization and power allocation problem is non-convex, finding the global optimal solution is extremely difficult\cite{relaytc,Convex}. Therefore, it is desirable to achieve a suboptimal solution with an acceptable complexity. Based on the results in Section III-A and III-B, an efficient algorithm that can obtain suboptimal solution is designed. Since lower bounds are used to obtain a sequence of non-decreasing solutions, no global optimality can be guaranteed for our proposed algorithm.

As shown in Algorithm \ref{alg3}, the key idea of the proposed algorithm is to alternately optimize the transmit power and the trajectory. In each iteration, the main complexity of the proposed algorithm lies in the steps 3 and 6, which require solving a series of convex problems. The computational costs of steps 3 and 6 are about $O\left( {{{\left( {MN} \right)}^3}} \right)$ and $O\left( {{{\left( {2M} \right)}^3}} \right)$, respectively, where $M$ and $N$ are the numbers of time slots and nodes.
\begin{algorithm}[t]
\caption{Joint transmit power and trajectory optimization}
\label{alg3}
\begin{algorithmic}[1]
\small
\STATE Initialize the UAV's trajectory $\left\{ {{x}\left[ m \right],{y}\left[ m \right]} \right\}^l$ and iteration number $l=0$
\STATE \textbf{Repeat}
\STATE ~~Solve the problem (\ref{P11}) with given trajectory $\left\{ {{x}\left[ m \right],{y}\left[ m \right]} \right\}^l$ by standard convex optimization techniques
\STATE ~~Update the transmit power ${\left\{ {{p_m}\left[ n \right]} \right\}}^{l+1}$ and minimum average throughput $s^{l+1}$
\STATE ~~\textbf{Repeat}
\STATE ~~~~~Solve the problem (\ref{P22}) with given transmit power ${\left\{ {{p_m}\left[ n \right]} \right\}}^{l+1}$ and get the optimal solution ${\left\{ {\Delta _x^k\left[ m \right],\Delta _y^k\left[ m \right]} \right\}}$ at the $k$-th iteration
\STATE ~~~~~Update the trajectory ${x^{k + 1}}\left[ m \right] = {x^k}\left[ m \right] + \Delta _x^k\left[ m \right]$ and ${y^{k + 1}}\left[ m \right] = {y^k}\left[ m \right] + \Delta _y^k\left[ m \right]$
\STATE ~~\textbf{Until} $s^{k+1}-s^k \leq \varepsilon$
\STATE ~~Update the trajectory $\left\{ {{x}\left[ m \right],{y}\left[ m \right]} \right\}^{l+1} = \left\{ {{x}\left[ m \right],{y}\left[ m \right]} \right\}^k$
\STATE \textbf{Until} $s^{l+1}-s^l \leq \varepsilon$
\STATE Return the trajectory $\left\{ {{x^\ast}\left[ m \right],{y^\ast}\left[ m \right]} \right\}$ and transmit power ${\left\{ {{p_n}^ * \left[ m \right]} \right\}}$
\end{algorithmic}
\end{algorithm}
\section{Simulations and Discussions}
In this section, simulations are conducted to demonstrate the effectiveness of the proposed algorithm. Consider a $2000 \times 500$ $\text{m}^2$ area where a UAV provides wireless connectivity for three nodes with locations of $(200,400,0)$, $(1000,200,0)$ and $(1800,400,0)$, respectively. The unit bandwidth is considered and other system parameters are as follows: $\sigma ^2 = -169$ dBm/Hz, $H=100$ m, $V=100$ m/s, $T=50$ s, ${P_T}=5$ W and $\varepsilon=0.01$. Without loss of generality, the time slot length is chosen to be $\delta = 1$ s and thus the number of discrete points is $M=50$. The channel power gain at $d_0 = 1$ m is ${{\beta _0}}=10^{-3}$. Two scenarios are investigated, where the initial locations are both $(0,0,100)$ in two cases and the final locations are $(2000,0,100)$ in case I and $(2000,500,100)$ in case II, respectively. For the benchmark, we consider the case that the UAV flies from the initial location to the final location along a straight line at an uniform speed. This trajectory is also used as the initial trajectory for the algorithm 1. Moreover, a static access point placed in the geometric center of the ground nodes is also considered to demonstrate the benefit brought by the UAV's mobility.
\begin{figure*}
\centering
\subfigure[Trajectory in case I]{
\label{TrajectoryX}
\includegraphics[width=0.3\textwidth]{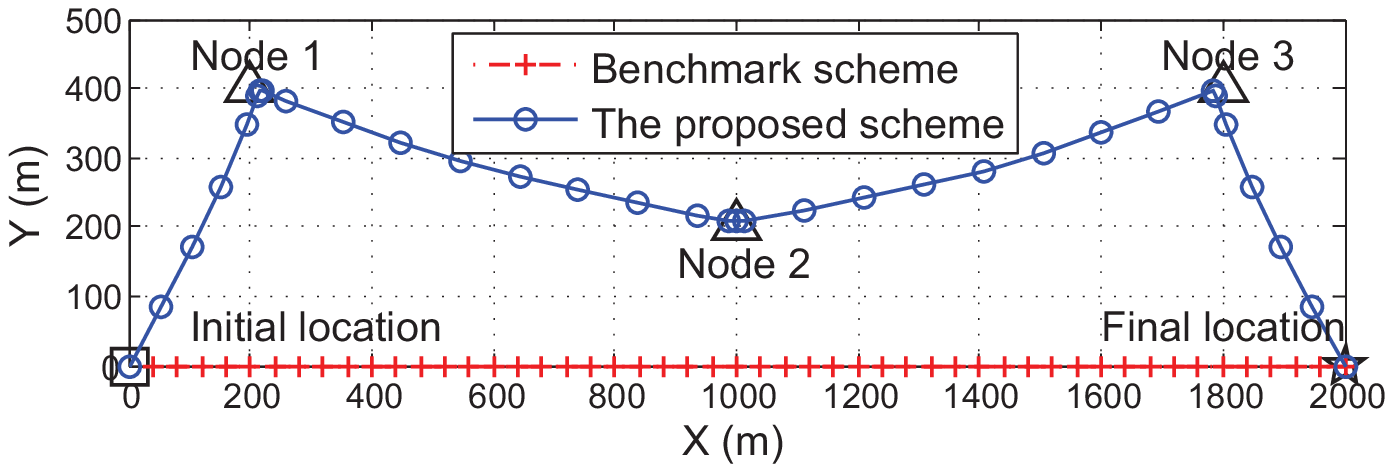}
}
\subfigure[Speed in case I]{
\label{TrajectoryX}
\includegraphics[width=0.3\textwidth]{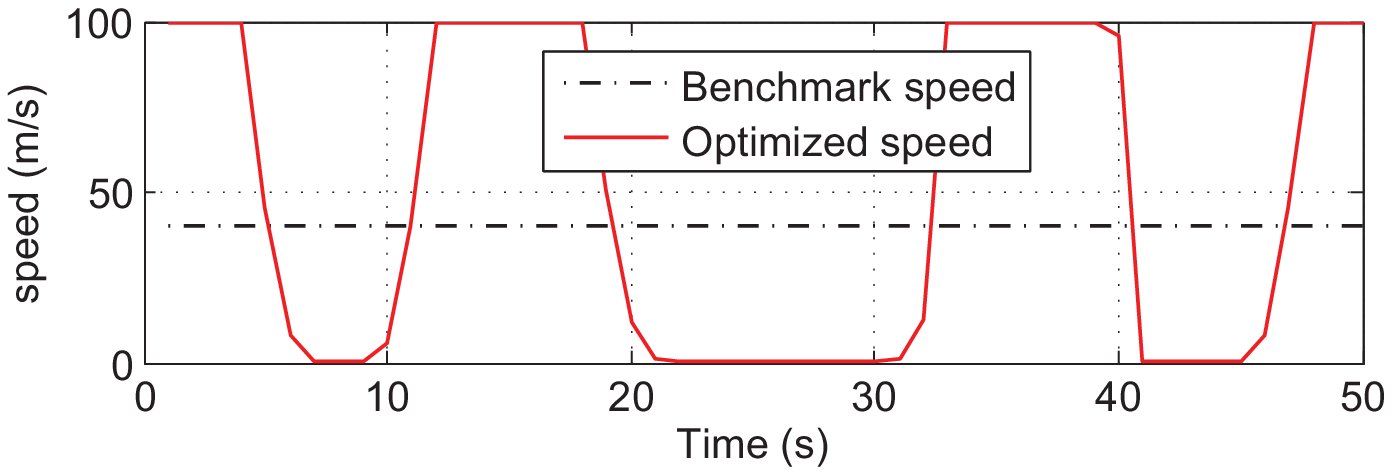}
}
\subfigure[Transmit power in case I]{
\label{TrajectoryX}
\includegraphics[width=0.3\textwidth]{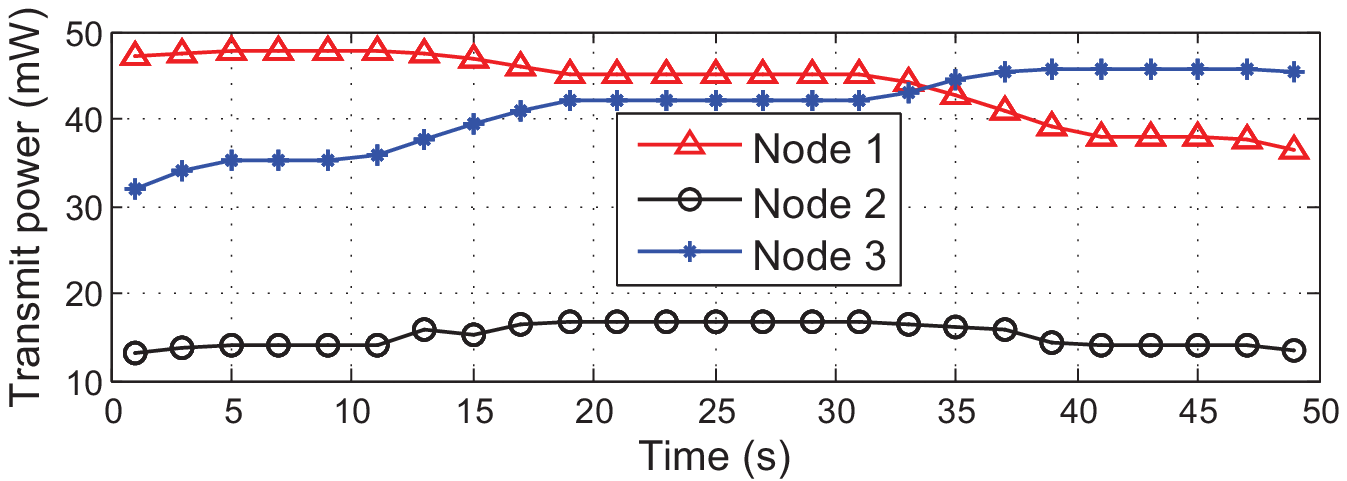}
}
\subfigure[Trajectory in case II]{
\label{TrajectoryY}
\includegraphics[width=0.3\textwidth]{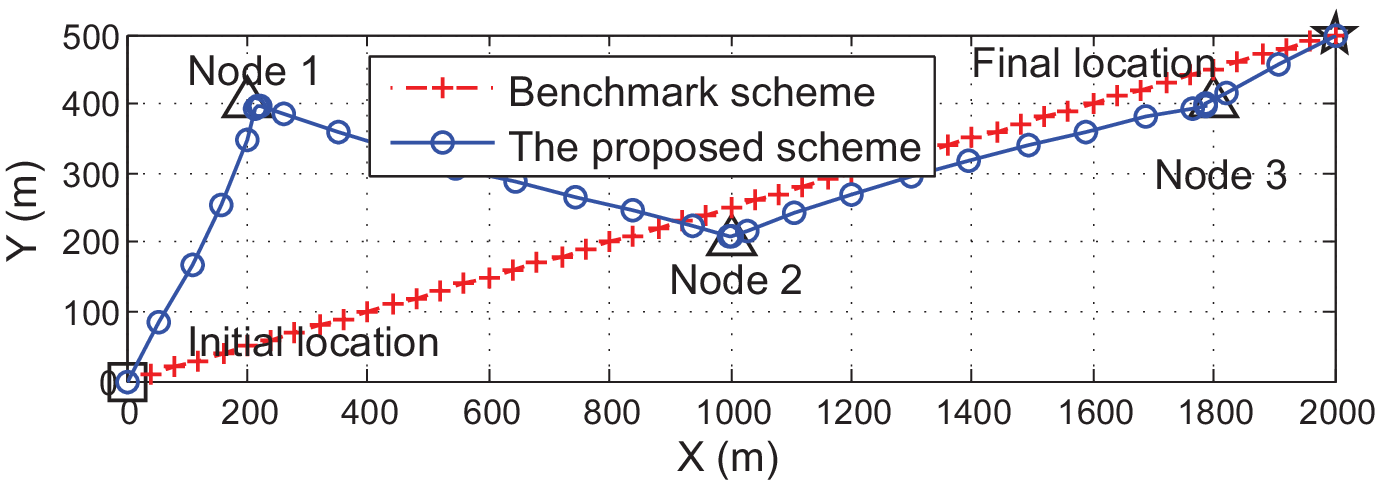}
}
\subfigure[Speed in case II]{
\label{TrajectoryY}
\includegraphics[width=0.3\textwidth]{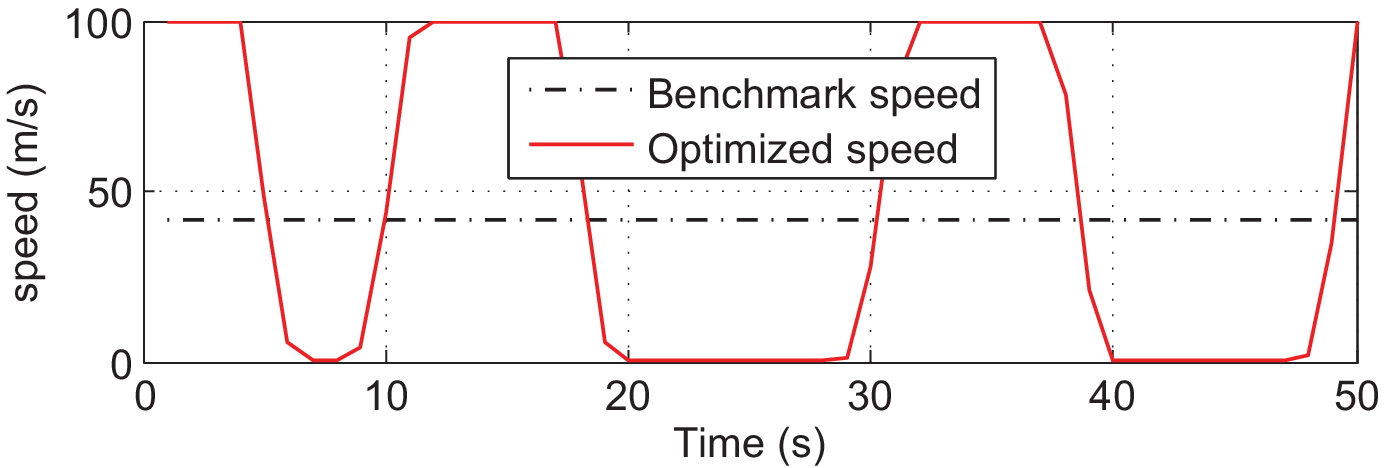}
}
\subfigure[Transmit power in case II]{
\label{TrajectoryY}
\includegraphics[width=0.3\textwidth]{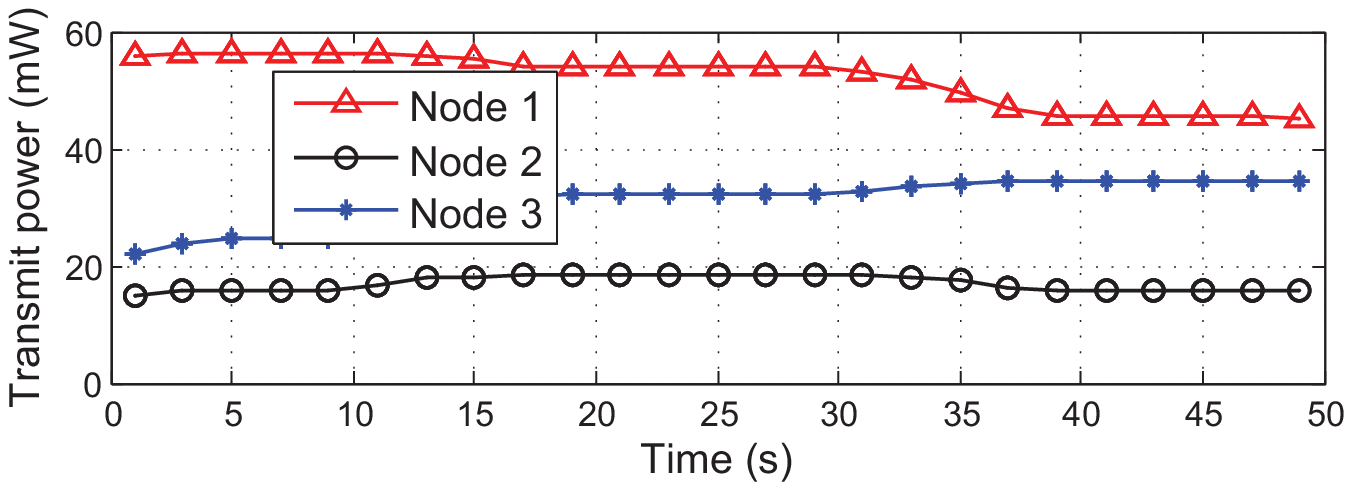}
}
\caption{The UAV's trajectory, speed and transmit power in considered scenes, where triangle, square and star represent the node, initial and final locations.}
 \label{fig1}
\end{figure*}

Fig. 1 presents the UAV's trajectory, speed and transmit power in two considered scenes. It can be observed that the optimized trajectory visits any nodes in both cases. In this case, the shortest trajectory is expected to provide more time that can be spent on hovering over the nodes. Moreover, the UAV's speed approaches 0 m/s at some points as shown in Fig. 1(b) and (e), which means that the UAV will hover over the nodes for a period of time.

Further, it can be observed from Fig. 1(c) and (f) that the transmit power is tightly related to the UAV's locations, which implies the necessity of joint transmit power allocation and trajectory optimization. The transmit power for node 2 is always lower than other nodes. This is because the UAV hovers a longer time over the node 2, as can be seen in Fig. 1(b) and (e). In addition, a phenomenon similar with water-filling can be observed in spatial domain. Specifically, the transmit power will be higher when the UAV approaches the node, which means better channel state. Conversely, when the UAV is away from the node, the corresponding transmit power becomes lower.

\begin{figure}[!t]
\centering{\includegraphics[width=80mm]{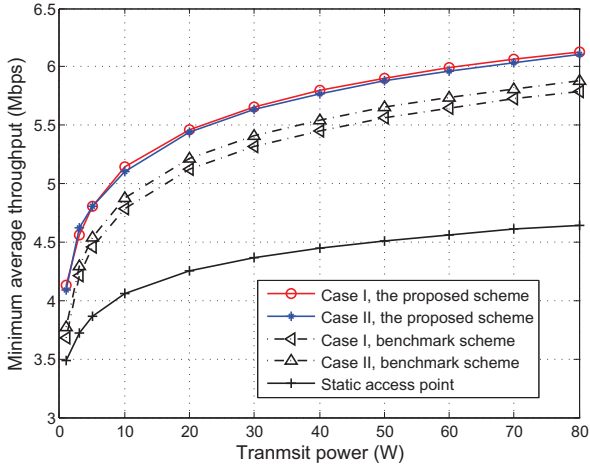}}
\caption{The achievable minimum average throughput versus the transmit power.}
\label{throughput}
\end{figure}
To evaluate the performance of the proposed algorithm, the minimum average throughput versus the transmit power is investigated as shown in Fig. \ref{throughput}. It is observed that the proposed algorithm outperforms the benchmark method and static case. The main reason is that the optimized trajectory provides better link quality and the proposed algorithm concentrates most of the power to time slots with the best link qualities.
\section{Conclusion}
In this letter, transmit power and trajectory optimization problem for UAV-aided networks was investigated, where a UAV acting as a mobile access point provides network access for some wireless nodes. The UAV's trajectory and transmit power were jointly optimized to achieve max-min average throughput. Simulation results validated the superiority of the proposed algorithm and revealed that the transmit power shows a water-filling characteristic in spatial domain.

\ifCLASSOPTIONcaptionsoff
  \newpage
\fi

\end{document}